\documentclass[aps,prd,preprint,superscriptaddress,tightenlines,nofootinbib]{revtex4}
\newcommand{\PRE}[1]{{#1}}   

\newcommand{\zt}[1]{\rm{#1}}

\usepackage{amsfonts}
\usepackage{amsmath}
\usepackage{amssymb,amsbsy}
\usepackage{bm}
\usepackage{makeidx}
\usepackage{latexsym}
\usepackage{graphicx}
\usepackage{epsfig}
\usepackage{subfigure}
\usepackage{dsfont}
\usepackage{mathcomp}
\usepackage{color}
\usepackage{mathrsfs}

\newcommand{\postscript}[2]{\setlength{\epsfxsize}{#2\hsize}
   \centerline{\epsfbox{#1}}}
\newcommand{\comment}[1]{}

\begin{document}

\preprint{
\hfil
\begin{minipage}[t]{3in}
\begin{flushright}
ANL-HEP-PR-10-53\\
NUHEP-TH/10-23
\vspace*{.4in}
\end{flushright}
\end{minipage}
}

\title{Searching for the Layered Structure of Space at the LHC
\PRE{\vspace*{0.3in}}}

\author{Luis A. Anchordoqui}
\affiliation{Department of Physics,\\ University of Wisconsin-Milwaukee, Milwaukee, WI 53201, USA
\PRE{\vspace*{.1in}}
}

\author{De Chang Dai}
\affiliation{HEPCOS, Department of Physics,\\ SUNY at Buffalo, Buffalo, NY 14260-1500, USA
\PRE{\vspace*{.1in}}
}

\author{Haim \nolinebreak Goldberg}
\affiliation{Department of Physics,\\
Northeastern University, Boston, MA 02115, USA
\PRE{\vspace*{.1in}}
}

\author{Greg  Landsberg}
\affiliation{Department of Physics,\\  Brown University, Providence,
RI 02912, USA
\PRE{\vspace*{.1in}}
}

\author{Gabe Shaughnessy}
\affiliation{
High Energy Physics Division,\\  Argonne National Laboratory, Argonne, IL 60439, USA
\PRE{\vspace*{.1in}}
}
\affiliation{
Department of Physics and Astronomy,\\ Northwestern University, Evanston, IL 60208, USA
\PRE{\vspace*{.1in}}
}

\author{Dejan Stojkovic}
\affiliation{HEPCOS, Department of Physics,\\ SUNY at Buffalo, Buffalo, NY 14260-1500, USA
\PRE{\vspace*{.1in}}
}

\author{Thomas J. Weiler}
\affiliation{Department of Physics and Astronomy,\\
Vanderbilt University, Nashville, TN 37235, USA
\PRE{\vspace*{.1in}}
}

\date{December 2010}

\begin{abstract}
  \noindent Alignment of the main energy fluxes along a straight line
  in a target plane has been observed in families of cosmic ray
  particles detected in the Pamir mountains. The fraction of events
  with alignment is statistically significant for families with
  superhigh energies and large numbers of hadrons. This can be
  interpreted as evidence for coplanar hard-scattering of secondary
  hadrons produced in the early stages of the atmospheric cascade
  development. This phenomenon can be described within the recently
  proposed ``crystal world,'' with latticized and anisotropic spatial
  dimensions.  Planar events are expected to dominate particle
  collisions at a hard-scattering energy exceeding the scale
  $\Lambda_3$ at which space transitions from
  3D~$\rightleftharpoons$~2D.  We study specific collider signatures
  that will test this hypothesis.  We show that the energy-spectrum of
  Drell-Yan scattering is significantly modified in this framework. At the LHC, two jet and
  three jet events are necessarily planar, but four jet events can
  test the hypothesis. Accordingly, we study in a model-independent
  way the $5\sigma$ discovery reach of the ATLAS and CMS experiments
  for identifying four jets coplanarities.  For the extreme scenario
  in which all $pp \to 4~\zt{jet}$ scattering processes become
  coplanar above $\Lambda_3$, we show that with an integrated
  luminosity of 10(100) fb$^{-1}$ the LHC experiments have the
  potential to discover correlations between jets if $\Lambda_3 \alt
  1.25(1.6)$~TeV.  \end{abstract}

\maketitle

\section{Introduction}

An intriguing alignment of gamma-hadron families (i.e., the outgoing
high energy secondary particles from a single collision in the
atmosphere) along a straight line in a target (transverse) plane has
been observed with (lead and carbon) $X$-ray emulsion chambers
($X$REC's) in the Pamir mountains~\cite{:1986et}.\footnote{The
  Pb-chambers are assembled of many sheets of lead (1~cm thick)
  interlaid with $X$-ray films. This provides a few interaction
  lengths for hadrons and a quasicalorimeter determination of the
  particle's energy. The C-chambers contain a 60~cm carbon layer
  covered on both sides by lead plates sandwiched with $X$-ray
  films. The carbon block provides a large cross section for hadron
  interaction, while the lead blocks are of minimal thickness allowing
  determination of particle energies. The total area of the chambers
  is few tens of square meters. Electron-photon cascades initiated by
  high energy hadrons and gamma-rays inside the $X$REC's produced dark
  spots whose sizes are proportional to the cascade energy deposited
  on the $X$-ray film.}  These families can be reconstructed by
measuring the coordinates and the incident direction of each particle
in the film emulsion. This allows determination of the total energy in
gamma-rays and the total energy of hadrons release to
gamma-rays. Recall that most of the hadrons in the family are pions
and the average fraction of energy transferred by pions to the
electromagnetic component is $\simeq 1/3$.  All families in the
experiment are classified by the value of the total energy observed in
gamma-rays, $\sum E_\gamma$. The centers of the main energy fluxes
deposited on the $X$-ray film (a.k.a. ``subcores'') include halos of
electromagnetic origin, gamma-ray clusters, single gamma-rays of high
energy, and high energy hadrons. The criterion for alignment is given
by the asymmetry parameter
\begin{equation}
\lambda_N = \frac{1}{N (N-1) (N-2)}\sum_{i\neq j \neq k} \cos 2 \varphi_{ij}^k \,,
\end{equation}
where $N$ is the number of subcores and $\varphi_{ij}^k$ is the angle
between vectors issuing from the $k$-th subcore to the $i$-th and
$j$-th subcores~\cite{Ivanenko:1992qw}. The parameter $\lambda_N$
decreases from $1$ (corresponding to $N$ subcores disposed along a
straight line) to $-1/(N-1)$ (corresponding to the isotropic case).
Events are referred to as aligned if the $N$ most energetic subcores
satisfy $\lambda_N \geq \lambda_N^{\rm cut}$. A common choice is $N=4$
and $\lambda_N^{\rm cut} = 0.8$.

The data have been collected at an altitude of 4400~m a.s.l., {\it
  i.e.}, at a depth of 594~g/cm$^2$ in the atmosphere. For low energy
showers, $30~{\rm TeV} \alt \sum E_\gamma \alt 200~{\rm TeV}$, the
fraction of aligned events coincides with background expectation from
fluctuations in cosmic ray cascade developments. However, for $\sum
E_\gamma > 700~{\rm TeV}$, the alignment phenomenon appears to be
statistically significant~\cite{Mukhamedshin:2005nr}.  Namely, the
fraction ($f$) of aligned events is $f(\lambda_4 \geq 0.8) = 0.43 \pm
0.17$ (6 out of 14) in the Pb-$X$REC catalogue, and $f(\lambda_4 \geq
0.8) = 0.22 \pm 0.05$ (13 out of 59) in the C-$X$REC catalogue. The
predominant part of the gamma-hadron families is produced by hadrons
with energy $E_0 \agt 10 \sum E_\gamma$, corresponding to interactions
with a center-of-mass energy $\sqrt{s} \agt 4$~TeV.  Data analyses
suggest that the production of most aligned groups occurs low above
the chamber~\cite{Ivanenko:1992qw}. Thus, it is not completely
surprising that the KASCADE Collaboration has found no evidence of
this intricate phenomenon at sea level ($\sim
1000$~g/cm$^2$)~\cite{Antoni:2005ce}.

Interestingly, the fraction of events with alignment registered in
Fe-$X$REC's at Mt. Kanbala (in China) is also unexpectedly
large~\cite{Xue:1999bb}. For gamma-hadron families with energy $\sum
E_\gamma \geq 500$~TeV the fraction of aligned events is $f(\lambda_3
\geq 0.8) = 0.5 \pm 0.3$ (3 out of 6). In addition, two events with
$\sum E_\gamma \geq 1000$~TeV have been observed in stratospheric
experiments~\cite{Capdevielle:1988pe}. Both events are highly aligned:
{\it (i)} the so-called STRANA superfamily, detected by an emulsion
chamber on board a Russian stratospheric balloon, has $\lambda_4 =
0.99$; {\it (ii)} the JF2af2 superfamily, detected by an emulsion
chamber during a high-altitude flight of the supersonic aircraft
Concord, has $\lambda_4 = 0.998$. It is worth noting that
stratospheric experiments record the alignment of particles, whereas
mountain-based facilities register the alignment of the main fluxes
of energy originated by these particles on a target plane.

The strong collinearity of shower cores has been interpreted as a
tendency for coplanar scattering and quasiscaling spectrum of
secondary particles in the fragmentation
region~\cite{Deile:2010mv}. If the aligned phenomenon observed in
cosmic ray showers is not a statistical fluctuation, then events with
unusual topology may be produced at the Large Hadron Collider (LHC).
In this paper we carry out a systematic study of $pp \to 4~\zt{jet} $
scattering processes to establish the sensitivity of LHC experiments
to such planar-shape topology.

The analysis technique described herein constitute an entirely general
approach to search for planar scattering at the LHC. Firstly, we
generate Standard Model (SM) QCD events that contain 4 light jets and 2
$b$-quark jets and 2 light jets utilizing the ALPGEN Monte Carlo
code~\cite{Mangano:2002ea}. Next, the event shape variable is
classified according to the standard aplanarity parameter,
$A_p$~\cite{Barger:1987nn}. Namely, we define the signal region as
$A_p < A_p^{\rm cut}$ and the control region with $A_p > A_p^{\rm
  cut}$, varying the cut among these three choices: $A_p^{\rm cut} =
0.1, 0.05, 0.01.$ After that, we extract the ratio of bi-planar to
planar events ($N_b/N_p$) and calculate the $N_b/N_p$ uncertainty
based on Poisson statistics for a given luminosity $L$. Finally, we
calculate the required luminosity to obtain $5\sigma$, $3\sigma$, and
95\% C.L. away from various values of $N_b/N_p$.  Though the search
technique is agnostic regarding the hypothetical physics underpinning
the planar configuration of events, an observation could have bearing
on the recently proposed idea that spatial dimensions collapse at
short distance, shutting off one-by-one with rising
energy~\cite{Anchordoqui:2010er}.

The layout of the paper is as follows. In Sec.~\ref{IV} we study the LHC sensitivity to coplanar particle escape.  We quantify signal and background rates of 4 jet events and show that future measurements of the aplanarity distribution of multi-jet events can provide a potent method for exposing a dimensional reduction of phase space. In Sec.~\ref{II} we associate the dimensional reduction in momentum space to a reduction in spatial dimensions.  We assume that space at its fundamental level is an anisotropic lattice~\cite{Anchordoqui:2010er}. This idea that spatial dimensions effectively reduce with increasing energy directly constrasts with field/string theories in continuous spacetime dimensions, where dimensionality increases with a rise in energy.  In Sec.~\ref{III} we present a phenomenological analysis of the Drell-Yan scattering processes~\cite{Drell:1970wh} for a 3D~$\rightleftharpoons$~2D crossover and we show that Tevatron  data is insufficient to constrain the model for a dimensional reduction above 1~TeV.  We summarize our conclusions in Sec.~\ref{V}.

\section{Sensitivity of the LHC experiments to 4 planar-jet events}
\label{IV}

In this section we estimate the LHC sensitivity to coplanar events  and corresponding discovery reach at $\sqrt{s}=14~\text{ TeV}$ using $2 \to 4$ scattering processes.  Such scattering processes involve multiple virtual particles. 
We define $\Lambda_3$ as the energy-scale of the onset of new physics.
We assume that when the momentum transfer $Q$ ($Q^2 = - \hat t$) in each of the propagators is comparable with $\Lambda_3$, 
a growing fraction of the jets are produced in one plane in their center-of-mass frame. This coplanarity is drastically different from the usual topology of 4D scattering, where the four outgoing partons are in general acoplanar. 
For simplicity, we consider three values of the coplanarity fraction at $\Lambda_3$.
We investigate the fractions 30\%, 50\%, and 100\%, utilizing simulated QCD events (with ALPGEN~\cite{Mangano:2002ea}) containing 4 jets (either 4 light jets or 2 $b$-quark jets and 2 light jets). While a fraction of the simulated events is planar, the entire sample is generally bi-planar.  Given the limited number of events containing four high transverse momentum ($p_T$) jets, the probe of truly planar events is statistics limited.  Therefore, we look to compare the SM QCD prediction of bi-planar jets with a purely planar sample by determining how many events are required for the observation of {\it aplanar} events to be significant.

For a given propagator scale, $\Lambda_3$, we accept events that have 4 jets that pass the following acceptance cuts:
\begin{equation}
p_{T,j} \ge \frac{1}{2}\Lambda_3 \text{ GeV}, \quad \eta_j < 2.5, \quad \Delta R_{j,j} > 0.4,
\end{equation}
where $\eta_j$ is the $j$ jet pseudorapidity, and $\Delta R_{j,j}$ is the separation in the azimuthal angle ($\phi$) - pseudorapidity ($\eta$) plane among jets:
\begin{equation}
\Delta R_{j1,j2} = \sqrt{ (\phi_{j1} - \phi_{j2})^2+(\eta_{j1} - \eta_{j2})^2}.
\end{equation}
In addition, to reject events that do not have a hard momentum scale, $Q\sim \Lambda_3$, we require that the invariant mass of any two pairs of jets satisfy
\begin{equation}
M_{jj} > \Lambda_3,
\end{equation}
and that for any two pairs of jets, the transverse momentum of a jet relative to the boost axis of the jet pair is
\begin{equation}
k_T> \frac{\Lambda_3}{2}.
\end{equation}
Finally, we model detector resolution effects by smearing the final state jet energy according to:
\begin{equation}
\frac{\Delta E}{E} = \frac{0.50}{\sqrt{E/\text{GeV}}} \oplus 0.03.
\end{equation}

\begin{figure}[t]
\begin{minipage}[t]{0.49\textwidth}
\postscript{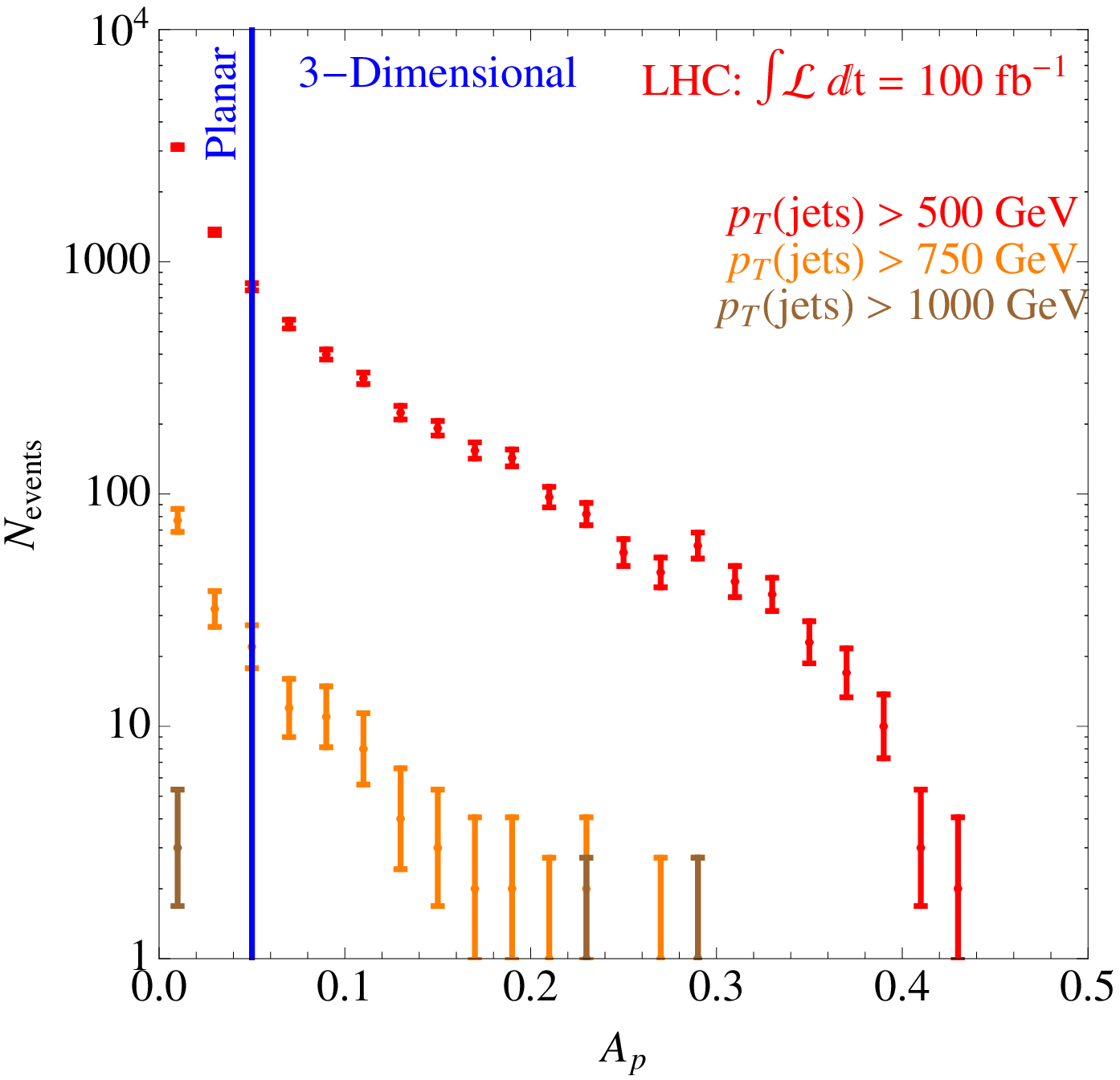}{0.9}
\end{minipage}
\hfill
\begin{minipage}[t]{0.49\textwidth}
\postscript{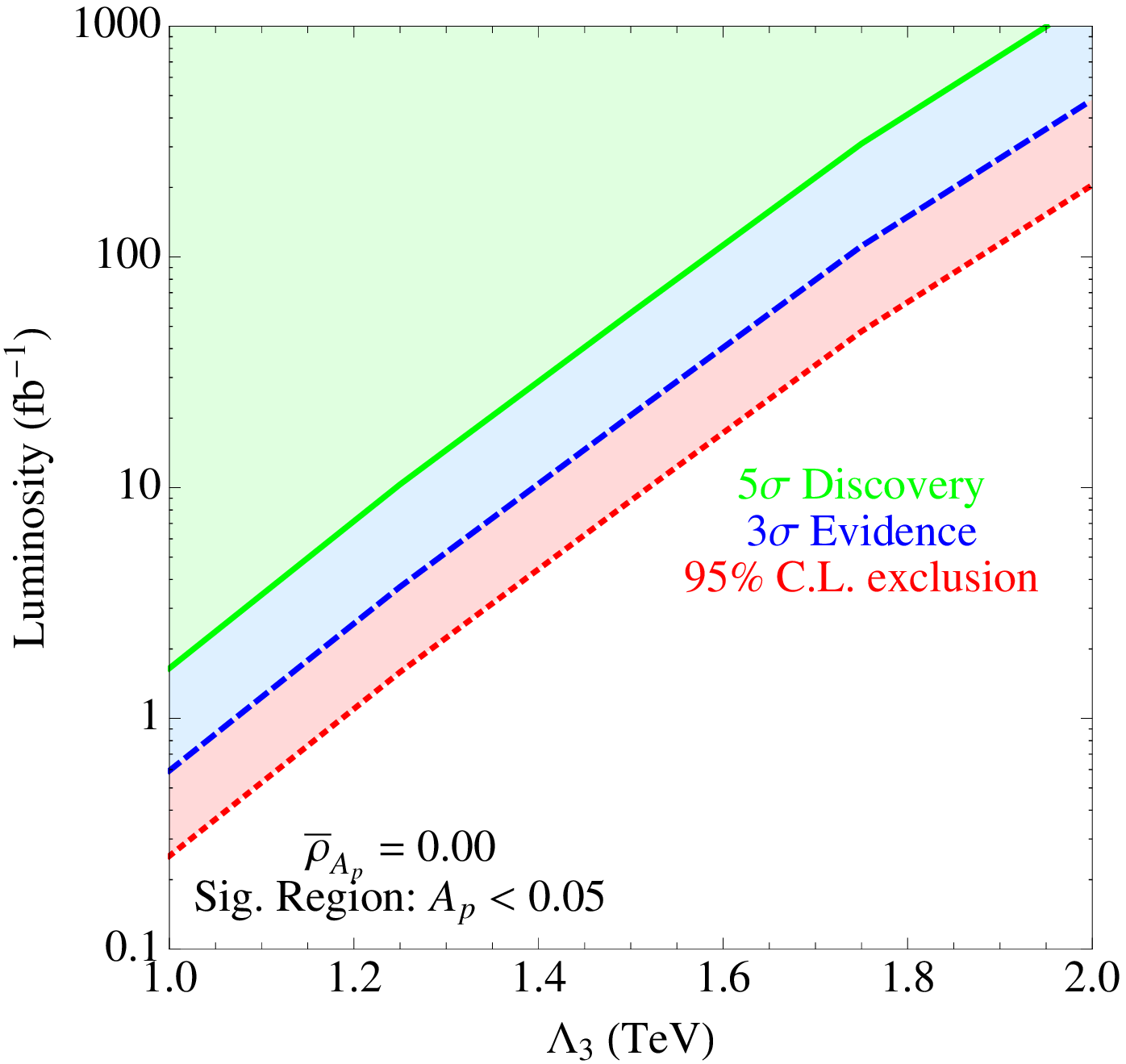}{0.9}
\end{minipage}
\begin{minipage}[t]{0.49\textwidth}
\postscript{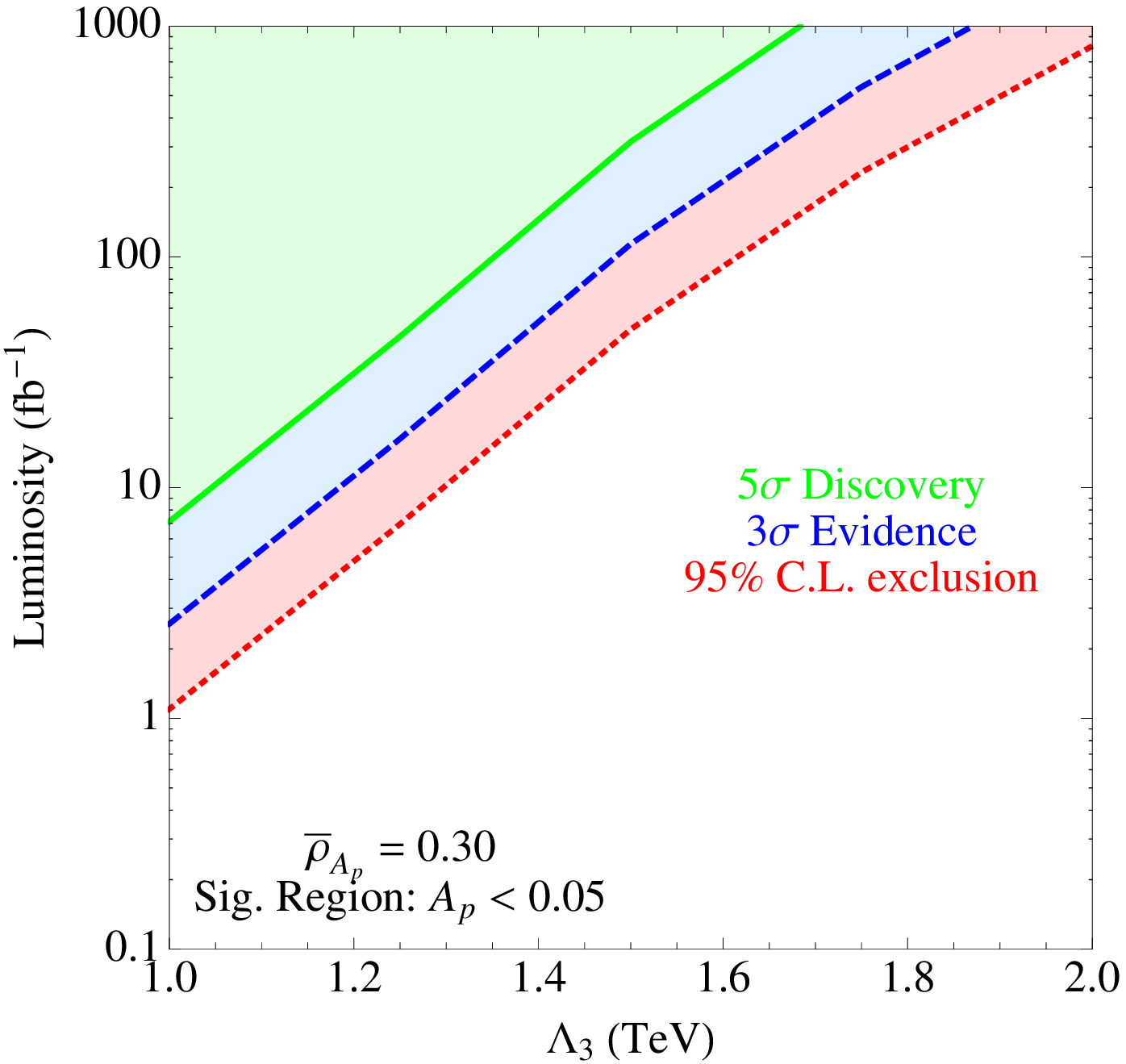}{0.9}
\end{minipage}
\hfill
\begin{minipage}[t]{0.49\textwidth}
\postscript{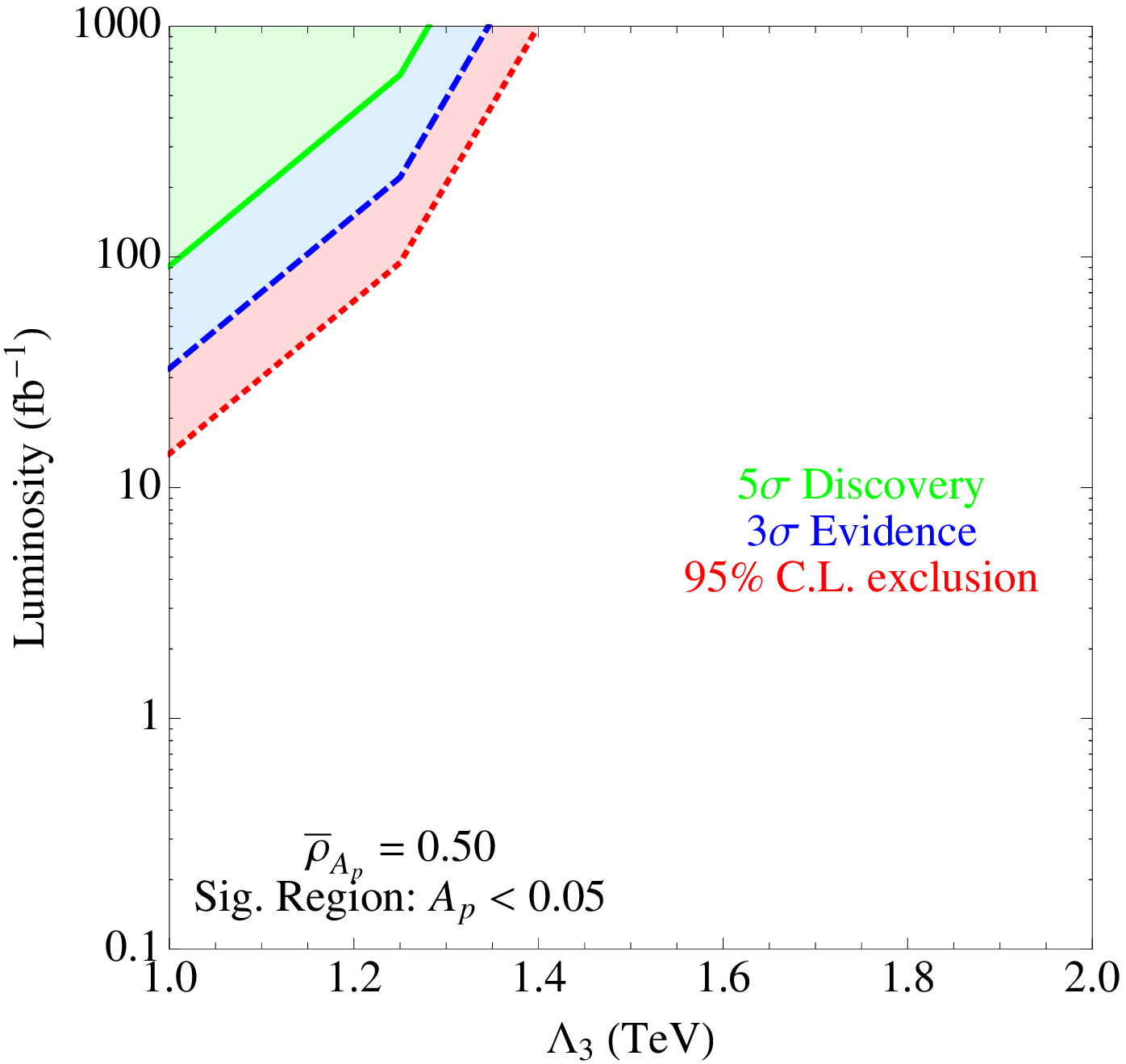}{0.9}
\end{minipage}
\caption{{\it (a)} Aplanarity of SM QCD 4-jet events for 100~fb$^{-1}$ and jet $p_T$ acceptance cuts (from above) of 500, 750 and 1000 GeV.   We take events with $A_p < 0.05$ as being planar. {\it (b)} Reach of $\Lambda_3$ at the LHC based on the aplanarity event shape variable. The $5\sigma$ discovery is indicated by a solid line, the $3\sigma$ evidence by a dashed line, and the 95\% CL exclusion by a dotted line.}
\label{fig:1}
\end{figure}

The event shape variable, aplanarity, can be calculated based on the constructed momentum tensor
\begin{equation}
M_{ab} = \frac{\sum_i k_{ia} k_{ib}}{\sum_i k_i^2}
\end{equation}
where $i$ runs over the 4 jets in each event, with all momenta measured  in the c.m.
The aplanarity is defined as~\cite{Barger:1987nn}
\begin{equation}
A_p = \frac{3}{2} Q_1,
\end{equation}
where $Q_1$ is the smallest normalized eigenvalue of the momentum tensor, giving $A_p$ a maximum value of $\frac{1}{2}$.  Therefore, planar or collinear events possess $A_p\sim 0$ values, while more 3D events approach the maximum value, $A_p= \frac{1}{2}$.

Now, we define events as exhibiting a planar topology by having the aplanarity below some cut value, $A_p < A_p^{\rm cut}$.  In Fig.~\ref{fig:1}.a, we show $A_p$ for a variety of jet $p_T$ cuts for the LHC with 100 fb$^{-1}$ of integrated luminosity.  To facilitate distinguishing planar events from the usual SM  QCD bi-planar events, we define the variable
\begin{equation}
\rho_{\cal O}=N_b / N_p,
\end{equation} the ratio of 3D bi-planar events to planar events using the observable ${\cal O}$ as a discriminator for planar and bi-planar events.  For pure planar events, this ratio should vanish, while for 3D SM  events, it is generally nonzero.  For a given $p_T$ cut, we can identify at what luminosity it is statistically different from a ratio the model predicts, i.e. $\bar \rho_{\cal O}$.  We vary the value of $A_p^{\rm cut}$ and find that $A_p^{\rm cut}=0.05$ gives the maximum sensitivity for $\bar \rho_{A_p}=0$, denoted by the vertical line in Fig.~\ref{fig:1}.a.  In Fig.~\ref{fig:1}.(b-d), we show the reach the LHC may have for a given luminosity for planar events, recast into the propagator scale, $\Lambda_3$ for values of $\bar \rho_{A_p}$.  With 10(100) fb$^{-1}$, the LHC may discover propagators of scale $\Lambda_3\approx 1.25(1.6)$ TeV with $\bar \rho_{A_p}=0$.  The reach is degraded as $\bar \rho_{A_p}$ increases such that at $\bar \rho_{A_p}=0.5$, $5\sigma$ discovery at 100 fb$^{-1}$ may be possible only for $\Lambda_3=1$ TeV.  This strong dependence on $\bar \rho_{A_p}$ is due to the 3D region in Fig.~\ref{fig:1}a having fewer events than the planar region.

\begin{figure}[t]
\begin{minipage}[t]{0.49\textwidth}
\postscript{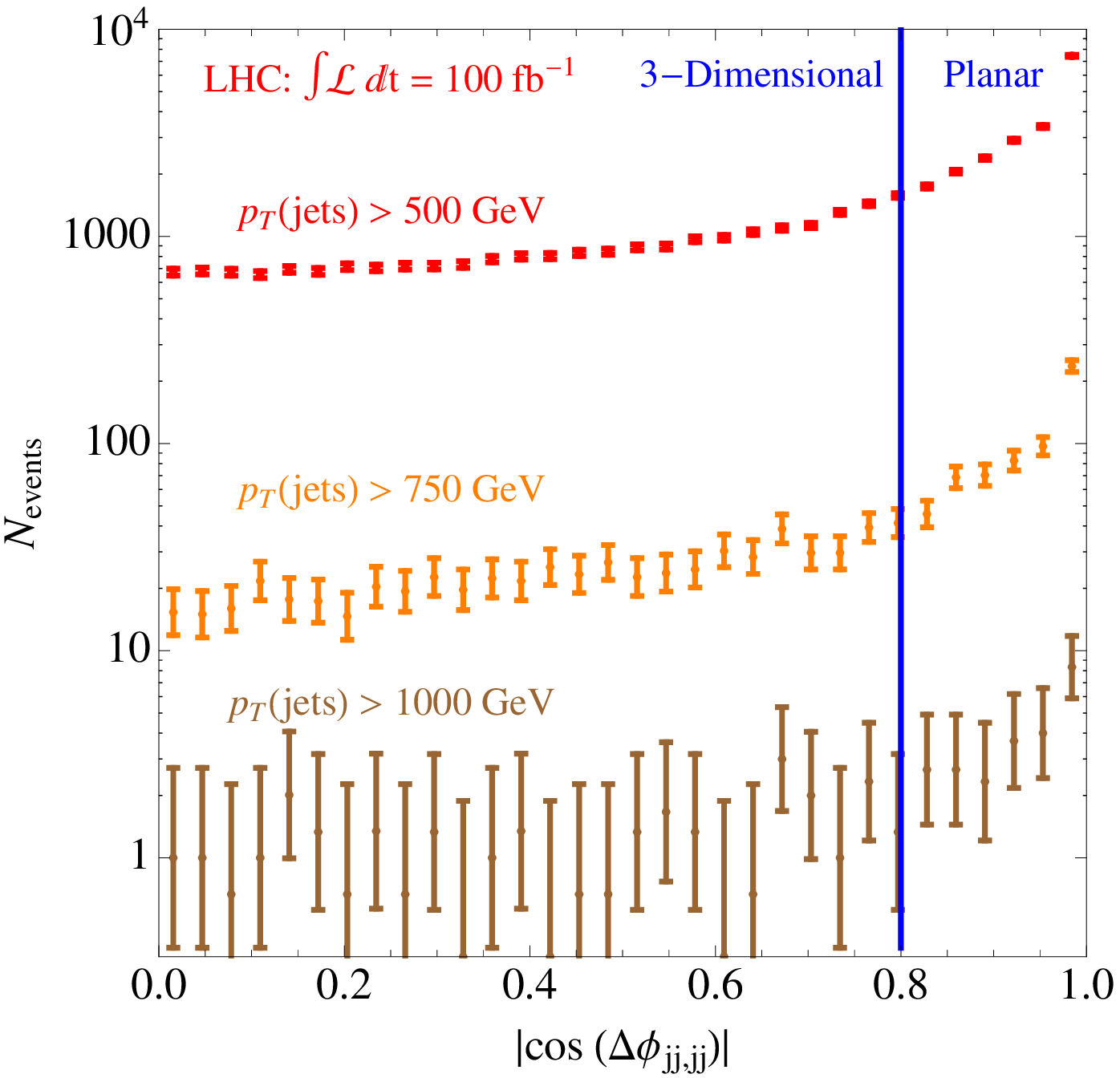}{0.9}
\end{minipage}
\hfill
\begin{minipage}[t]{0.49\textwidth}
\postscript{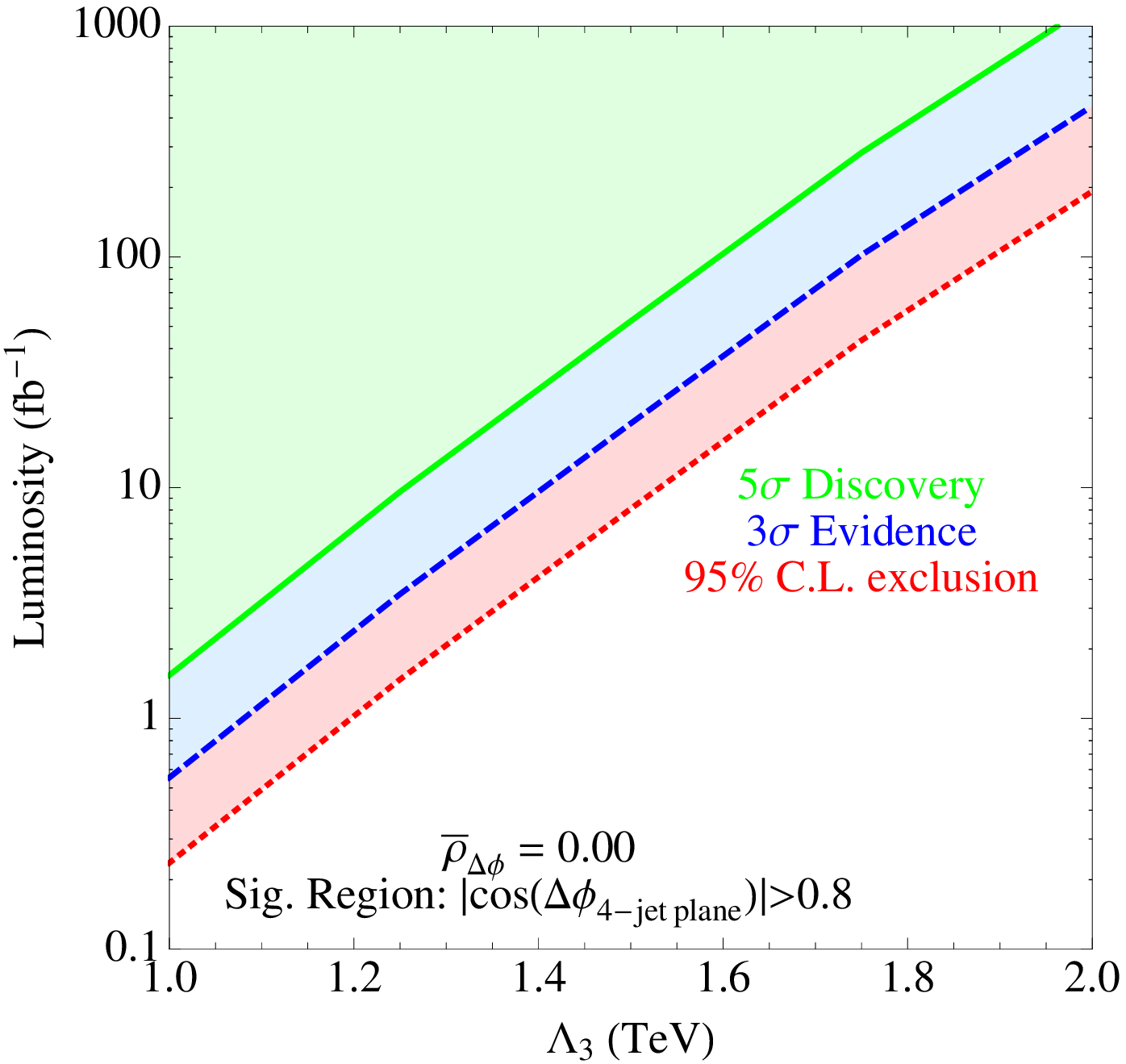}{0.9}
\end{minipage}
\begin{minipage}[t]{0.49\textwidth}
\postscript{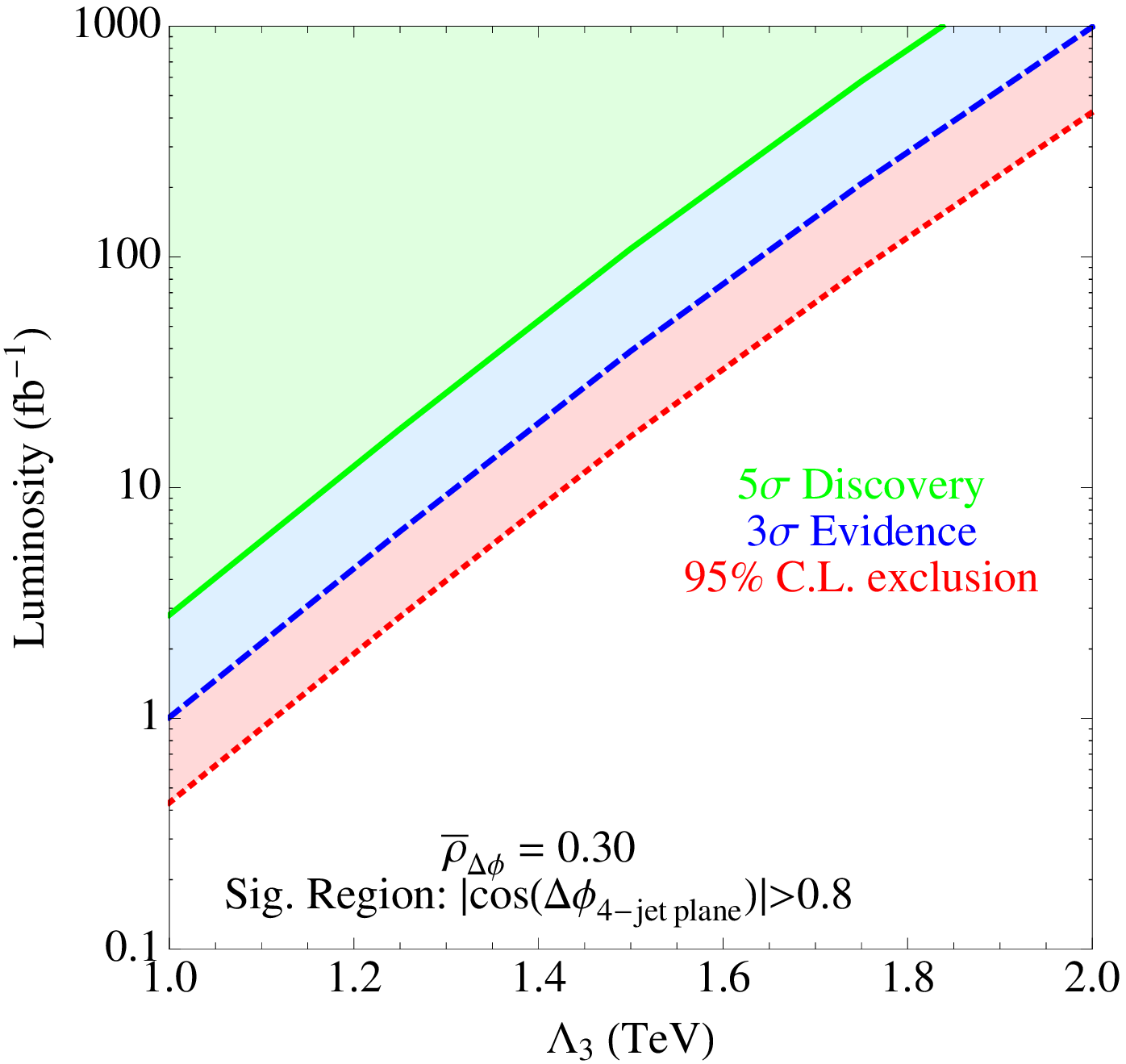}{0.9}
\end{minipage}
\hfill
\begin{minipage}[t]{0.49\textwidth}
\postscript{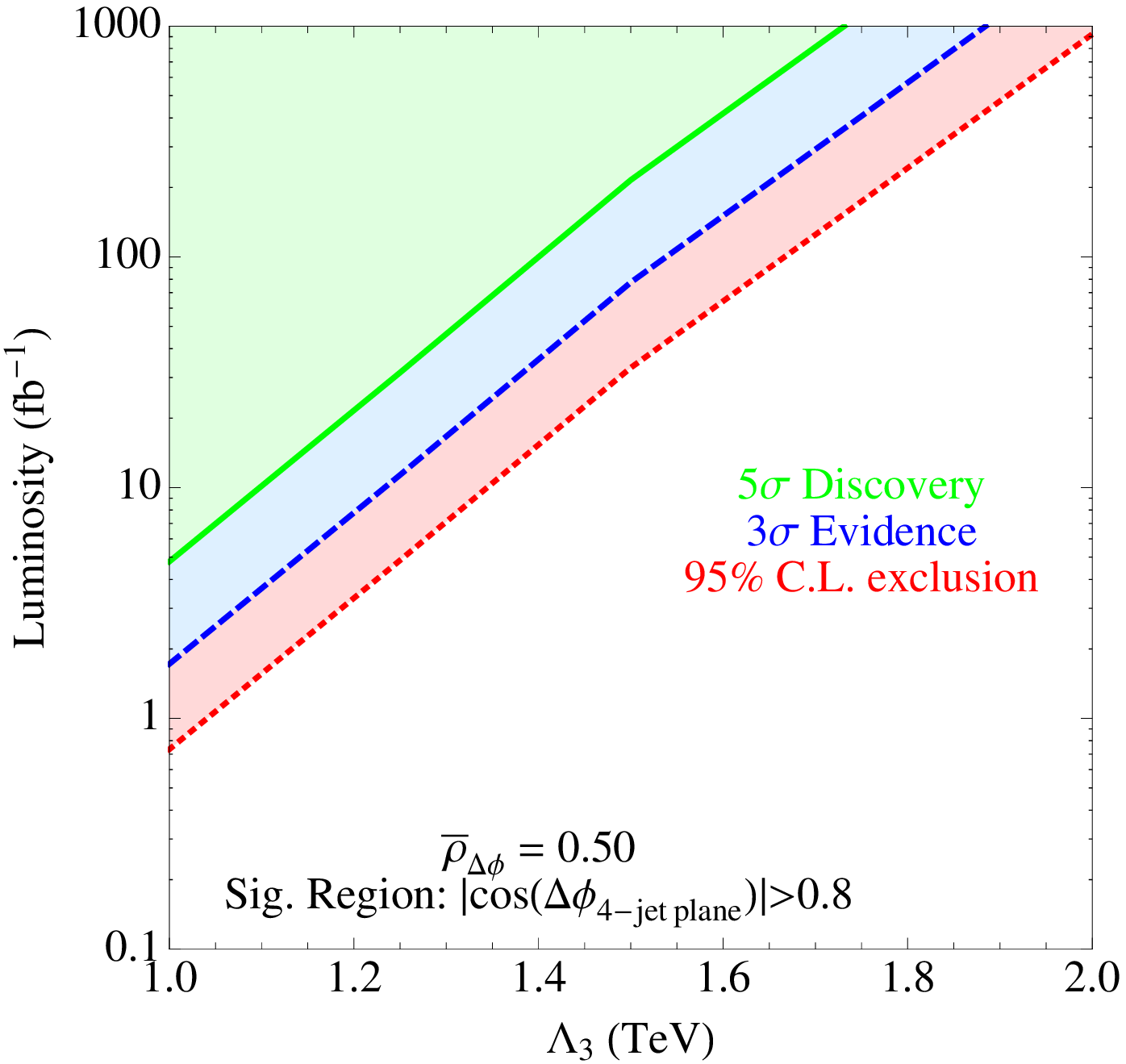}{0.9}
\end{minipage}
\caption{{\it (a)} The $|\cos(\Delta \phi_{ jj,jj})|$ for 100~fb$^{-1}$ and jet $p_T$ acceptance cuts (from above) of 500, 750 and 1000~GeV,  where $\Delta \phi_{ jj,jj} \equiv \Delta \phi_{\rm 4-jet \, plane}$ is the angle between the two planes defined by the respective jets combinations. There are a total of 3 combinations, and each are included. We define the signal region as $|\cos (\Delta \phi_{\rm 4-jet \, plane})| > 0.8.$ {\it (b)} Reach of $\Lambda_3$ at the LHC based on $|\cos (\Delta \phi_{\rm 4-jet \, plane})|$. The $5\sigma$ discovery is indicated by a solid line, the $3\sigma$ evidence by a dashed line, and the 95\% CL exclusion by a dotted line.}
\label{fig:2}
\end{figure}

In addition to the aplanarity, one can construct other variables
to test for planar events.  For instance, two planes may be defined
as going through each jet pair.  The subsequent azimuthal angle,
$\Delta \phi_{ jj,jj}$, between these two planes should be zero
for all combinations of jet pairings for planar events, and nonzero
otherwise.  When using this alternate variable and placing a cut as
done for the $A_p$ analysis above (see Fig.~\ref{fig:2}.a), we arrive
at a very similar reach for $\bar \rho_{\Delta \phi}=0$, shown in
Fig.~\ref{fig:2}.b; other $\bar \rho_{\Delta \phi} \neq0$ cases are
shown in Fig. 3(c,d).  The weaker dependence of the reach on the
nonzero value of  $\bar \rho_{\Delta \phi}$ than in the $A_p$ case
is due to the smaller $\rho_{A_p}$ value in the SM.  Thus, making it
easier to statistically distinguish between the bi-planar SM and planar
events.\footnote{Note, however, that the values
$\bar \rho_{A_p}$ and $ \bar \rho_{\Delta \phi}$
are not necessarily equivalent.}

An approach for  production of highly coplanar multi-jet events, which  exploits semi-hard QCD processes with a high transverse momentum transfer, has been discussed in~\cite{Halzen:1989rg}. 
This model is susscesful in explaining coplanar 3-jet events in cosmic ray data. The leading particle provide one jet and the collinear singularity of QCD correlates the second and third jets, therefore producing a roughly coplanar 3-jet events in the lab system. The enhanced amplitude due to collinear gluon emission may explain, for example, the two jets ``ridge'' phenomenon recently observed at the LHC in $pp$ collisions~\cite{Khachatryan:2010gv} and previouly observed at RHIC in heavy ion collisions\cite{Adams:2005dq}. However, QCD collinearity cannot explain $N\geq 4$ jet events in cosmic ray physics, which corresponds to $N \geq 3$ jet events in the center-of-mass system (which is also the lab system) at the LHC. In the remainder of this paper we will discuss a model where multijet events are coplanar for all values of $N$.

\section{The crystal world}
\label{II}

\begin{figure}[htb]
\scalebox{0.5}{\includegraphics{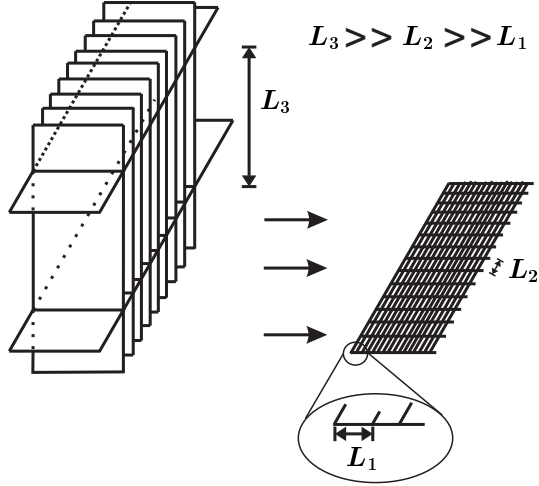}}
\vspace*{-0.1in}
\caption{Ordered lattice. The  fundamental quantization scale of space  is indicated by $L_1$. Space structure is $1D$ on scales
  much shorter than $L_2$, while it appears
  effectively $2D$ on scales much larger than $L_2$ but
  much shorter than $L_3 $. At scales much larger than
  $L_3$, the structure appears effectively $3D$~\cite{Anchordoqui:2010er}.}
\label{lattice}
\end{figure}

Motivated by condensed matter systems, some us recently proposed spacetime may be an ordered lattice structure that becomes anisotropic at very small distances~\cite{Anchordoqui:2010er}. The proposed set up, shown in Fig.~\ref{lattice}, resembles that of dimensional crossover in layered strongly correlated metals~\cite{Valla}. These materials have an insulating character in the direction perpendicular to the layers at high temperatures but become metal-like at low temperatures, whereas transport parallel to the layers remains metallic over the whole temperature range.  The analogy which we adopt is to replace the temperature variable in the materials system with short-distance ``virtuality'' in the parton scattering processes.  We further assume that the lattice orientation is randomized on a scale sufficiently small to avoid any preferred direction in space on the macroscopic scale.  On the small scale, there will be a preferred direction given by the local lattice orientation.  Therefore, hard scattering processes can resolve the lattice spacings that separate ``conducting space'' from the ``crystal world'' of insulating space.  On the other hand ``macroscopic'' objects like beam protons effectively see a spacetime continuum.

 It is  of interest to explore the consequences of this extreme viewpoint (and let experiment be the arbiter). We are not aware of any data that would rule out this conjecture.  We emphasize that this conjecture is radically different from the increase of dimensions at small spatial scales that are postulated in string theories and in many modern field theories. In contrast with our conjecture the effective number of dimensions decreases as partons probe smaller scales.  With the reduction in spatial dimension, phase space is reduced, the cross-sections are reduced, and multi-jet final states 
are necessarily coplanar (for $3\to  2$).  With an increase in spatial dimension, phase space is increased, cross-sections increase, and multi-jet final states fill the three spatial dimensions but also lose energy and multiplicity into the extra dimensions.

As demonstrated in Sec.~\ref{IV} for the coplanarity, striking collider signals would be observed at the LHC if $\Lambda_3 \sim 1/L_3 \sim 1$~TeV.  Many related aspects of beyond the SM field theory will occur at high-energy: the transition of the renormalizable SM to a super-renormalizable field theory, modification in the evolution properties of parton distribution functions, running coupling constants, running anomalous dimensions of operators, and so on.  A study of these new effects requires a more specific model.  Such a study is beyond the scope of this paper.  If the LHC provides indications for the correctness of our conjecture, then it becomes neccessary and even imperative to explore the new physics in detail.

In the next section we address an obvious probe of new physics, namely Drell-Yan production of lepton pairs.  Above $\sqrt{s} \sim \Lambda_3$, the reduction of phase space is expected to reduce the Drell-yan cross-section, We show that present Tevatron data does not impact on the conjecture, but that future LHC data will provide a new reach in $\Lambda_3$.

\section{Drell-Yan meets the Crystal World}
\label{III}

In this section we assess the effect on Drell-Yan cross sections at
Tevatron and at LHC of the restriction to coplanarity of scattering
above the lattice energy $\sim \Lambda_3\sim$ 1 TeV. In order to
effectuate this, we will need to hypothesize an algorithm for a smooth
continuation between $(2+1)D$ to $(3+1)D$.

In a standard manner, we define the cross section as the probability transition rate divided by the flux,
and  obtain
\begin{eqnarray}
d \sigma_{_D} & = & \frac{1}{(2\pi)^{D-2}}  \frac{1}{16 \hat s} \frac{p_f^*}{p_i^*}
 \left|{\cal M}_D \right|^2 \, d \Omega_{D-1}^* |\vec p_f^*|^{D-4} \nonumber \\
 & = & \left( \frac{\left|{\cal M}_D \right|}{8 \pi \sqrt{\hat s}} \right)^2
 \left(\frac{p_f^*}{p_i^*} \right) d \Omega_{D-1}^* \left( \frac{|\vec p_f^*|}{2 \pi} \right)^{D-4} \, ,
\label{eq:eqqD}
\end{eqnarray}
where ${\cal M}_D$ is the  Feynman amplitude invariant in $D$ dimensions;
 $p_i^*$ and $p_f^*$ are the initial and final center-of-mass momenta $\sim  \sqrt{\hat s}/2$; and
$d \Omega_{D-1}^*$ is the solid angle aperture in the center of mass.
Thus
\begin{equation}
\sigma_4  \propto  {\cal M}_4 ^2/\hat s,
\end{equation}
 whereas for  $D=3$
\begin{equation}
 \sigma_3  \propto  {\cal M}_3 ^2/\hat s^{3/2}  \, .
\label{eq:34}
\end{equation}
(Note that since ${\cal M}_3\sim E^1,$  the
``cross section'' $\sigma_3$ in two spatial dimensions is a straight line with
dimension  length, as expected.)

To assess the impact on Drell-Yan, it is of critical importance to modify $\sigma_3$ in
order that it gain entry as a $\sigma_4$ in the parton model
calculation of $pp(\bar p) -> \ell^+\ell^- + X$. This entails a change in
dimension for the cross sections, and thus it {\em cannot} occur
in the probability transition rate, which has the same dimension $(E^1)$ for any $D$.
Thus it must reside in the incoming flux~\cite{Calmet:2010vp}.  Assuming that this
modification is purely of geometric origin, without energy dependence, we
insert a factor of $\Lambda_3$ to adjust dimensions and obtain the following
prescription for the transition to $D=3$:
\begin{eqnarray}
\label{eq:fourthree}
\sigma_4^{\rm eff}  &\sim& \sigma_3/\Lambda_3 \nonumber \\
&\sim & ({\cal M}_3 ^2/\Lambda_3)\ / \hat s^{3/2} \end{eqnarray}

At this point we must deal with the matrix element ${\cal M}_3$. There are essentially two
choices: {\it (i)} one can assume total ignorance of the dynamics,
(seeing as it may involve the interaction of particles with a space
time lattice), and simply work on dimensional grounds. This approach is decidedly unsatisfactory, since
it would permit arbitrary powers of $\sqrt{\hat s}/\Lambda_3,$ vitiating any hope of making contact
with quantitative experimental findings;
or {\it (ii)} we can assume that ${\cal M}_3$ can be calculated
from a 3 dimensional version of QED. In case {\it (ii)}, the operational Lagrangian is
\begin{equation}
{\cal L}_{\rm QED_3} = i\bar\psi\gamma^\mu(\partial_\mu - ie_3A_\mu)\psi -
\tfrac{1}{4}F^{\mu\nu}F_{\mu\nu} \,,
\label{eq:lag}
\end{equation}
where $F_{\mu \nu} = \partial_\mu A_\nu - \partial_\nu A_\mu$ is the field strength, $e_3$ is the coupling, and $\psi$ is a four-component spinor with the corresponding four-dimensional representation of the Dirac algebra
\begin{equation}
\gamma^0 = \left(\begin{array}{cc} \sigma^3 & 0 \\ 0  & -\sigma^3 \end{array}\right) \, ,  \quad \quad
\gamma^1 = \left(\begin{array}{cc} i \sigma^1 & 0 \\ 0 & - i \sigma^1 \end{array}\right), \quad \quad
\gamma^2 = \left(\begin{array}{cc} i \sigma^2 & 0 \\ 0 & - i\sigma^2 \end{array}\right) \, ,
\end{equation}
with $\sigma^i$ the Pauli matrices~\cite{Jackiw:1980kv}.

Then the calculation of Drell-Yan  proceeds the same manner as
in $D=4$, with the result
\begin{equation}
{\cal M}_3 \sim e_3^2\ \hat s/Q^2\sim e_3^2  .
\label{eq:em3}
\end{equation}
where the factor $\hat s$ comes from the Dirac spinor normalization
and the $Q^2$ from the virtual photon (or $Z$) propagator. An
engineering dimensional analysis of (\ref{eq:lag}) shows $ e_3^2\sim
E^1,$ so that indeed ${\cal M}_3 \sim e_3^2 \sim {\rm const} \sim
\Lambda_3$.\footnote{In a more general situation $e_3$ would be energy
  (and dimension) dependent, receiving contributions from radiative
  corrections~\cite{Shirkov:2010sh}.}  In this case,
Eq.(\ref{eq:fourthree}) reads
\begin{equation}
\sigma_4^{\rm eff}  \sim \Lambda_3/ \hat s^{3/2}
\label{eq:seff}
\end{equation}
for $\sqrt{\hat s} > \Lambda_3.$ More explicitly, the working algorithm for
examining effects of the transition to
coplanarity would be the following:
\begin{eqnarray}
\sigma^{\rm DY} &=& \sigma^{\rm DY}_{\rm SM}~~~~~~~~~~~~~~~~~~~~ \sqrt{\hat s} \ll \Lambda_3 \nonumber \\
                 &=&\sigma^{\rm DY}_{\rm SM} \ (\Lambda_3/\sqrt{\hat s})
                 ~~~~~~~~\sqrt{\hat s} \gg \Lambda_3 \, ,
                 \label{eq:working}
\end{eqnarray}
where $\sigma^{\rm DY}_{\rm SM}$ is the Drell-Yan parton-parton cross section as calculated in the SM. For convinience we use as an interpolating function
\begin{equation}
\sigma^{\rm DY} = \sigma^{\rm DY}_{\rm SM} \,  \left(1 - e^{-\Lambda_3/\sqrt{\hat s}}\right) \, ,
\label{interpolation}
\end{equation}
where $e^{-\Lambda_3/\sqrt{\hat s}}$ parameterizes the the probability to remain confined to the two dimensional space.  The quantity $\Lambda_3$ may also be regarded as characteristic of a potential barrier for phase transitions from 3 to 2 spatial dimensions, with $e^{-\Lambda_3/\sqrt{\hat s}}$ representing the probability of tunneling between these states. This is perhaps reminiscent of the the factor $\sim {\cal E}^2\; e^{-\pi m^2/e |{\cal E}|}$ the probability per unit time per unit volume for creating an $e^+e^-$ pair in a constant electric field ${\cal E}$~\cite{Schwinger:1951nm}. To obtain the total cross section for the process $pp(\bar p) -> \ell^+\ell^- + X$ we have to convolute (\ref{eq:working}) with the parton distribution functions.

We do not at present have a controlled calculation coming from a
well-defined formalism for the parton distribution with particle
momenta oblique to the lattice layer. A crude approximation that
conserves 4-momentum among the particles is to assume the beam axis is
aligned with the lattice layer. This hadron alignment in turn means
that the partons in the infinite momentum frame are aligned with the
lattice. In reality, it is only a projection of the parton momentum
that is aligned with the lattice.  Each of the two
partons oblique to the lattice will presumably undergo an inelastic
scattering with the lattice plane in advance of their mutual hard
scattering.\footnote{The single parton is assumed to coherently,
  elastically scatter from the lattice, thereby altering its refractive
  index (velocity) in a small way.}  This new mechanism of parton energy loss  is somewhat analogous to an initial state radiation off the partons beyond the physical radiation included in factorization theorems. 

A question that immediately arises is whether there are constraints on
this model from Tevatron data. The D0 Collaboration reported the most
recent study of the dielectron invariant mass spectrum analyzing
5.4~fb$^{-1}$ of data collected at $\sqrt{s} =
1.96$~TeV~\cite{Abazov:2010ti}. There are four events observed in the
energy bin $600~{\rm GeV} - 800~{\rm GeV}$. Since there is negligible
background from other SM processes, the 68.27\%CL spread in the
Poisson signal mean is ($2.34,\, 6.78$)
events~\cite{Feldman:1997qc}. At this point it is worth recalling that
the Drell-Yan $d\sigma^{\rm DY}/dQ$ measures lepton pair spectra at
parton collision energy $Q.$ Therefore, from Eq.~(\ref{interpolation})
it is straightforward to verify that a dimensional reduction at
$\Lambda_3 = 1$~TeV (which predicts $\simeq 3$ events in the $600~{\rm
  GeV} - 800~{\rm GeV}$ energy bin) is consistent with Tevatron data
at the $1\sigma$~level. A similar analysis follows from the invariant
mass spectrum of dimuon data reported by the CDF
Collaboration~\cite{Aaltonen:2008ah}. Interestingly, the $\Lambda_3
\sim 1$~TeV region will be tested by the early LHC run at $\sqrt{s} =
7$~TeV.

\section{Conclusions}
\label{V}

In the first part of this paper we have presented a complete model independent study to search for planar events in 4~jet final states. The only free parameter of our analysis is the characteristic energy scale for the onset of coplanarity, $\Lambda_3$.  For the extreme scenario in which all $pp \to 4~\zt{jet}$ scattering processes become coplanar above $\Lambda_3$, we have shown that with an integrated luminosity of 10(100) fb$^{-1}$ the LHC experiments have the potential to discover correlations between jets if $\Lambda_3 \alt 1.25(1.6)$~TeV. 

In the second part of this paper we have discussed some aspects of the recently proposed~\cite{Anchordoqui:2010er} latticized spatial dimensions, with different characteristic lattice spacings in each dimension.  We have studied specific collider signatures that will probe this idea.  In particular, we have shown that the predicted energy spectrum of Drell-Yan scattering is significantly modified in this model.  Remarkably, the anisotropic crystal world yields planar events when the energies of hard scatterings exceed the scale at which space transitions from 3D~$\rightleftharpoons$~2D. Therefore, four jet events at the LHC will exhibit striking planar alignment (if the parton-parton momentum transfer $Q$ exceeds the energy scale $\Lambda_3$ of the lattice). Jets with this strong azimuthal anisotropy may have been already observed by the Pamir Collaboration: the effect know as alignment, which cannot be explained by conventional physics.

\section*{Acknowledgments}
We would like to thank Malcolm Fairbairn for valuable
discussions. This work is partially supported by the US National
Science Foundation, under Grants No. PHY-0757598, PHY-0757959, and
PHY-0914893, US Department of Energy, under Grants
No. DE-FG02-91ER40688, DE-AC02-06CH11357, DE-FG02-91ER40684, and
DE-FG05-85ER40226, EU Marie Curie Network UniverseNet
(HPRN-CT-2006-035863), and the UWM Research Growth Initiative.


\end{document}